\documentclass[11pt]{article}

\usepackage{LobsterTwo}

\usepackage{lipsum} 
\def\rotatechartwo#1{\reflectbox{#1}}

\usepackage{bm}
\usepackage{floatrow} 
\pdfoutput=1
\usepackage{amsmath,amsfonts,amsthm}
\usepackage{esdiff}  
\usepackage{booktabs}  
\usepackage{url}  
\usepackage{hyperref}  

\usepackage[tableposition=bottom]{caption} 

\usepackage{cleveref}  
	\crefname{equation}{equation}{equations}
	\crefname{figure}{figure}{figures}	
	\crefname{table}{table}{tables}
\usepackage[skip=1.5pt,font=small]{caption}

\usepackage{bbm}

\usepackage[usenames,dvipsnames,svgnames,table]{xcolor}

\usepackage{graphicx}

\usepackage{tikz}  

\usepackage[sc]{mathpazo} 
\usepackage[T1]{fontenc} 
\usepackage{microtype} 

\usepackage{multicol} 

\usepackage[margin={1cm,1.5cm}]{geometry}
\usepackage{booktabs} 
\usepackage{float} 
\usepackage{hyperref} 

\usepackage{lettrine} 
\usepackage{paralist} 

\usepackage{abstract} 

\usepackage{titlesec} 
\renewcommand\thesection{\Roman{section}} 
\renewcommand\thesubsection{\Alph{subsection}} 
\titleformat{\section}[block]{\large\scshape\centering\bfseries}{\thesection.}{1em}{} 

\titleformat{\subsection}[block]{\scshape\centering}{\thesubsection.}{1em}{} 

\usepackage{fancyhdr} 
\DeclareCaptionFormat{myformat}{#1#2#3\hrulefill}
\usepackage{aurical}
\usepackage[T1]{fontenc}
\usepackage[utf8]{inputenc}

\usepackage{authblk}
\makeatletter
\title{\vspace{-15mm}\fontsize{13pt}{13pt}\selectfont\textbf{An Artificially-intelligent Means to Escape Discreetly from the Departmental Holiday Party\\\vspace{1.5mm}\textit{guide for the socially awkward}}} %
\author[1,2]{Eve Armstrong\thanks{evearmstrong.physics@gmail.com}\thanks{https://reality-aside.com/aprilfool}}
\affil[1]{Department of Physics, New York Institute of Technology, New York, NY 10023, USA}
\affil[2]{Department of Astrophysics, American Museum of Natural History, New York, NY 10024, USA}\par
\date{(Dated: April 1, 2020)}
\begin{document}
\maketitle 

\begin{abstract}
\noindent
We shall employ simulated annealing to identify the global solution of a dynamical model, to make a favorable impression upon colleagues at the departmental holiday party and then exit undetected as soon as possible.  The procedure, \lq\lq Gradual Freeze-out of an Optimal Estimation via Optimization of Parameter Quantification\rq\rq\ - GFOOEOPQ, is designed for the socially awkward.  The socially awkward among us possess little instinct for pulling off such a maneuver, and may benefit from a machine that can learn to do it for us.  The optimization rests upon Bayes' Theorem, where the probability of a future model state depends on current knowledge of the model.  Here, the model state vectors are party attendees, and the future event of interest is their disposition toward us at times following the party.  We want these dispositions to be favorable.  To this end, we first complete the requisite interactions for making favorable impressions, or at least ensuring that these people later remember having seen us there.  Then we identify the exit that minimizes the chance that anyone notes how early we high-tailed it.  Now, poorly-resolved estimates will correspond to degenerate solutions.  As noted, we possess no instinct to identify a global optimum all by ourselves.  This can have disastrous consequences.  For this reason, GFOOEOPQ employs an annealing procedure that iteratively homes in on the global optimum.  The method is illustrated via a simulated event taken to be hosted by someone in the physics department (I am not sure who), in a two-bedroom apartment on the fifth floor of an elevator building in Manhattan, with viable Exit parameters: front door, side door to a stairwell, fire escape, and a bathroom window that opens onto the fire escape.  Preliminary tests are reported at two real-life social celebrations.  The procedure is generalizable to corporate events and family gatherings.  Readers are encouraged to report novel applications of GFOOEOPQ, to help expand the algorithm.
\end{abstract}
\section{INTRODUCTION}
\begin{multicols}{2}
Try this: recall the last departmental social function you attended.  Details are dim in your memory, but you still suffer from the aftermath.  You had received the invitation with dread.  At the time you didn't know many people, and you figured you should make a good impression.  So you decided to steel yourself for 45 minutes and then sneak away.  This sounded easy in principle.  But as usual, you botched it. 

For the better part of an hour you stumbled through rote motions, vaguely hoping for the best.  You brought a carrot cake~\cite{carrotCake}, which nobody ate.  You made a well-rehearsed joke~\cite{joke}, which nobody got.  Your effort to appear relaxed left you sweat-stained and made that
\begin{figure}[H]
\centering
  \includegraphics[width=0.99\textwidth]{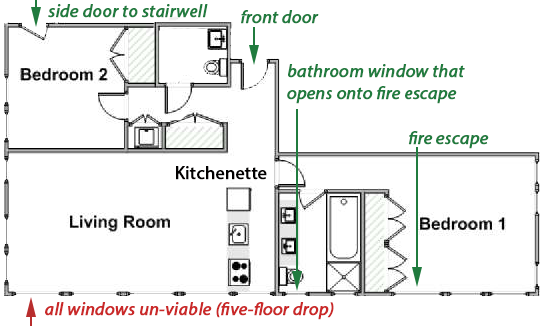}
  \caption{\textbf{Simulated geometry of state space, with exits labeled.}  Permitting reasonable risk, the green are viable and red is ruled out.}
  \label{fig:floorplan}
\end{figure}
\noindent
weird old eye tick flare up.  In short, you made a Herculean attempt to be fun and you failed anyway.  Finally, at the unjustifiably-early hour of 8pm, you tried to sneak out through a side door by the kitchen sink, caught your sweater on a splinter, and while flailing to free yourself ripped the sleeve in two whilst the spectacle was witnessed by the math professor's spouse's mistress~\cite{private1}, who told the math professor's spouse~\cite{private1,private2}, who told the math professor~\cite{private2,private3}, and come Monday the entire group was snickering about it over cookies at journal club~\cite{private3,private4,private5,private6,private7,private8,private9,private10,private11,private12,private13,private14,private15,private16,private17}. 

Has this happened to you?

If so\footnote{If not - that is, if you enjoyed the party because meeting people is fun - that's wonderful.  You need read no further.}, Dear Reader, GFOOEOPQ\footnote{Some have suggested that an alternate name be chosen for this procedure, although I do not see why.}  - Gradually Freezing Out an Optimal Estimation via Optimization for Parameter Quantification - can help.  GFOOEOPQ (g$\wedge$\textit{fu}i:\rotatechartwo{c}:p$\wedge$kw) is an inference blue-print, so to speak.  It is designed for those of us who - within a social setting - have no idea what to do~\cite{clegg2012stranger,sociallyAwkward}.  We do not recognize social cues, indeed perhaps are not even aware that social cues occur.  We survey the genuine friends whom we have acquired over the years, and marvel that we acquired them.  If you\footnote{If you are a student, attending these events does not matter much yet, but hone your skill early.  If you are tenured, you're fine; stay home.}, Reader, are a member of this blighted demographic, you are in good company.  GFOOEOPQ will permit you to formulate a social event on familiar terrain: one that is quantitative and optimizable.  The mere knowledge that you have this supportive system at hand will enable you to exercise rational thought - an ability that, in these situations, is often suppressed by panic.  Human instinct will not be necessary.  

GFOOEOPQ infers an estimation corresponding to the global solution of a dynamical model, where the state variables and unknown parameters of the model are the quantities to be estimated.  Once solved, the model may be integrated to, say, the Monday following a party.  The global solution is the set of estimates possessing the strongest predictive power regarding what will happen on Monday.  Much has been written on applications of inference to dynamical systems~\cite{kimura2002numerical,kalnay2003atmospheric,tarantola2005inverse,betts2010practical,crassidis2011optimal}.   For our purposes, the dynamical variables are features of the people at the party, some measurable and others hidden.  The parameters are relevant environmental characteristics, including viable modes of exit.  GFOOEOPQ estimates the optimal solution via the variational method to minimize a cost function~\cite{smith1998variational}.
  
Our objective is twofold.  First we seek to make positive impressions so that people recall us favorably later.  To this end, we need a measurable quantity, and we shall take it to be: facial expression, a common proxy for disposition~\cite{kaiser2001facial,horstmann2003facial,wang2009rapport}.  Our first task, then, is to enact the interactions that will maximize the likelihood of a positive facial expression upon a greeting on Monday.  Second, we aim to leave soon and unseen, lest being seen leaving soon negate the positive impression we have endeavoured to build.  In this phase, the critical parameter is the exit location.  Exit location will itself be a function of user-defined hyper-parameters, including densities of people in local regions of state space.

Now, this model is fiercely nonlinear and will yield multiple solutions.  One is optimal.  If we do not sufficiently resolve our options, we will fail to tell them apart.  Disaster may ensue.  For this reason, in each estimation phase the optimization is embedded within an annealing procedure~\cite{van1987simulated,ye2015systematic}, to freeze out the options as distinct. 

Armed with GFOOEOPQ, Reader, you will navigate your next social function with relative ease.  The algorithm has been tested extensively in simulations, wherein one can rewind and compare alternate strategies as many times as one pleases.  The simulated state space is a two-bedroom New York City apartment on the fifth floor of an elevator building, containing four viable exits (Figure~\ref{fig:floorplan}).  Four models were employed, each corresponding to a different constituency of people.  For each model, three sets of simulations were run, each seeking the global minimum corresponding to a particular future: a favorable disposition toward you, unfavorable disposition, and no memory of you whatsoever, respectively; these last two outcomes serve as instructive comparisons.  Preliminary results at two real-life events are also reported.

In \textit{Discussion} we shall explore a possible reverse formulation for the dynamical problem, wherein you sneak in late rather than leave early (only recommended if you have a spy inside).  Also noted is the severely sticky case in which you are the only one who shows up.  Finally, a caution to the socially-awkward reader: please mind your proclivity to taking statements literally.  While all efforts have been made in this manuscript to state the obvious, also peppered in are tongue-in-cheek comments, or: humor characterized by subtlety.  If you have questions regarding tone, please email for clarification.   

\textit{GFOOEOPQ is generalizable to industry and corporate events, graduations, baby's-first-birthdays, bridal showers, and wedding receptions.  I ask that GFOOEOPQ not be used at funerals, religious ceremonies, or any event catered by Zabar's~\cite{zabars}.  In such cases, please respectfully muster sincerity.}
\end{multicols}
\section{METHOD}
\begin{multicols}{2}

This Section walks you through a simulated evening at a physics department party purporting to celebrate an unspecified holiday.  Illustrations are taken from the simulations to be described in \textit{Results}.  

\subsection*{\textbf{0 \hspace{2mm} Define your model.}}

Before you arrive, define your foundation.  It will give you a sense of solid ground under your feet as you walk in the door.

Your model $\bm{f}$ is written in state vectors $\bm{x}^i$, where index $i = [1,2,...T]$ and $T$ is the number of people at this party.  Each vector is comprised of state variables $x_a^i$, or personal features; index $a = [1,2,...D]$ where $D$ is the number of features per person.  Some features, such as facial expression, are measureable.  Others, such as whether a person likes you, are not.  You would like to interact such that these people will like you come Monday.  Unless you have developed a procedure for mind-reading\footnote{Apparently someone has done this~\cite{domes2007oxytocin}.}, however, you cannot measure \lq\lq likes me\rq\rq; you need a measurable proxy.  You shall take that proxy to be facial expression.  In the optimization, you will use the information contained in facial expression during interactions to estimate all state variables - measurable and unmeasurable, and then predict the dynamics of facial expression upon a greeting on Monday. 

Each state variable $x_a^i$ evolves as: $\frac{d\bm{x}_a^i}{dt} = \bm{f}_a^i(\bm{x}^i)$.  For each state vector, $\bm{f}^i$ has three components: $\bm{f}_{intrinsic}^i$, $\bm{f}_{environment}^i$, and $\bm{f}_{me}^i$.  The first component, $\bm{f}_{intrinsic}^i$, are the dynamics intrinsic to Person $i$: the person's zeroth-order dynamics regardless of presence at the party.  You do not have much influence over this term, unless you, say, murder the person\footnote{Do not murder anybody.}, thereby setting $\bm{f}_{intrinsic}^i$ to zero.  The second term, $\bm{f}_{environment}^i$, is a first-order effect due to Person $i$'s presence within the environment of this party.  Unless you, say, set the kitchen on fire\footnote{Do not set the kitchen on fire.}, you do not control this term either.  The third term, $\bm{f}_{me}^i$, is your term.  These are the dynamics of your interaction with Person $i$.  You aim to tune these dynamics to maximize the likelihood that the component $x_a^i$ of state vector $\bm{x}^i$ that corresponds to Person $i$'s facial expression will be a welcoming smile upon an encounter come Monday.

Now, the components of $\bm{f}^i$ are coupled.  Person $i$'s intrinsic disposition will affect the manner in which that person interacts with you.  Thus, while you cannot control the first two terms of $\bm{f}^i$, you must study them to tailor $\bm{f}_{me}^i$ appropriately.  Note the $i$ index: your optimal means of operation on Person $i$ is not necessarily the optimal means for Person $j$.  If Katie likes being called Cupcake, that does not prove that Dean Carlos likes being called Cupcake.  

During the estimation window of GFOOEOPQ Phase One: \lq\lq Interact\rq\rq, the state vectors will also evolve according to unknown parameters $\bm{\theta}$.  The relative weights of the three model components are elements of $\bm{\theta}$.  If Cassandra laughs during your interaction, it could be because she thinks your joke is funny, or is enjoying the party, or simply laughs frequently.  As you interact, you will solidify your functional forms of $\bm{f}_{intrinsic}$, $\bm{f}_{environment}$, $\bm{f}_{me}$, and additional elements of $\bm{\theta}$\footnote{The form for $\bm{f}$, and definitions of features $x_a$ and $\bm{\theta}$ must be tailored to the event and to the user's aim regarding predictive power.  GFOOEOPQ currently supports all functions except those involving complex numbers and imaginary quantities.}.  

During the estimation window of GFOOEOPQ Phase Two: \lq\lq Exit\rq\rq, the parameters of interest become those related to exit location; let us denote these as hyper-parameters $\bm{\zeta}$.  These include the density of people within local neighborhoods of state space.  You will define these quantities as you sample the party geometry and atmosphere.  Do not worry about them for now; we shall examine them when it is time for you to vanish. 

\subsection*{\textbf{1 \hspace{2mm} Do you really have to go?}}

Before reading further, do yourself a favor.  Initialize $\bm{f}$, forward to Monday, and see what happens.  

Specifically, your priors $\bm{x}^i(0)$ for each person $i$ are all you already know about these people.  Start there.  Then run the state vectors through $\bm{f}$ for the party duration, applying only $\bm{f}_{intrinsic}$ and $\bm{f}_{environment}$ - leaving yourself out of the picture.  If the number of happy faces come Monday is already sufficient for your purposes, consider skipping this party.  

A caution, however: think strategically here.  Each event you attend buys you points out of attending some other event, which may be more or less unsavory than this one.  As a rule of thumb: if you don't anticipate this
\end{multicols}
\begin{figure}[H]
\centering
  \includegraphics[width=0.9\textwidth]{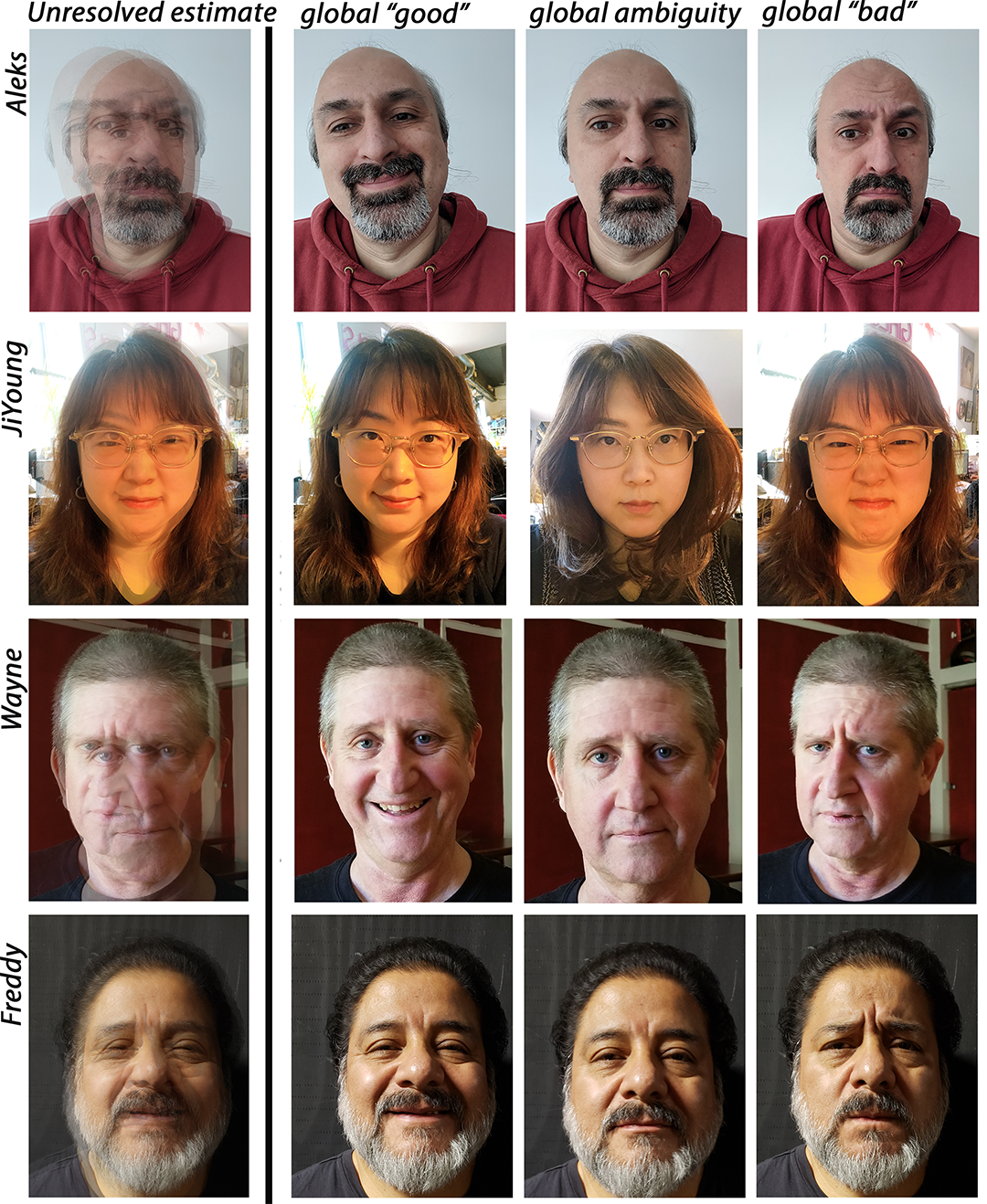}
  \caption{\textbf{Two-dimensional representation of state predictions for unresolved (Column 1) versus resolved (Columns 2-4) parameter estimates come Monday, illustrating the utility of annealing to avoid degenerate solutions.}  The coordinates are: \textit{identity} and \textit{facial expression}.  From top, the identities are: Aleks~\cite{aleks}, JiYoung~\cite{jiyoung}, Wayne~\cite{wayne}, and Freddy~\cite{freddy}.  At left (Column 1) is the prediction for each identity corresponding to simulations that did not employ annealing.  Note tip-offs for the user that indicate insufficient resolution; for example, Aleks has three pairs of eyes.  At right (Columns 2-4) are the predictions of three individual sets of simulations that employed annealing.  These sets correspond to the global solution for the \lq\lq good\rq\rq\ outcome (Column 2), and for comparison the \lq\lq ambiguous\rq\rq\ (Column 3) and \lq\lq bad\rq\rq\ (Column 4) outcomes.}
 \label{fig:faces}
\end{figure}
\begin{multicols}{2}
\noindent
one being all that bad, then go.  Win points for later.

\subsection*{\textbf{2 \hspace{2mm} Preliminaries}}

So you've opted in.  All right, here we go.  Preliminary do's and don'ts are as follows.

\subsubsection*{\textit{Do: install the algorithm on a hand-held device.}}

Lest you appear peculiar~\cite{weird}, perform these calculations discreetly.  Do not use the bathroom for this purpose.  Convergence can take time, and monopolizing the bathroom may make an undesirable impression.  Instead, run GFOOEOPQ in plain view, while appearing to chat on social media with a friend you've never met, or some other behavior that is deemed healthful.

\subsubsection*{\textit{Do Not: show up on time.}}

It is tempting to show up on the early side, to feel justified in leaving early.  This is not advised: it is a social faux pas, and it will not buy you time.  

True, the invitation states: \lq\lq 7pm - ???\rq\rq.  What the invitation \textit{means}, however, is to show up sometime after 7pm, certainly not before 7:10. The hosts will assume that this is understood and at 7pm might still be in the shower.

Any time after 7:15 is technically safe, but for the purposes of computation, wait\footnote{ Also, showing up at 7:15 will only be noted by the sparse few who are also there by 7:15.} until 7:30.  Initial conditions set at 7:15 are unlikely to represent the future state of the party.  Few state vectors are present yet at 7:15, so you will have to apply generous guesses.  As state vectors arrive, the model dimensionality will increase rapidly, and the space may undergo discontinuous transitions in mood and general atmosphere.  These factors might become hyper-parameters when it comes time to identify the optimal exit location.  For this reason, initializing your algorithm too early will at best waste time and at worst misguide the estimation and preclude convergence. 

\subsubsection*{\textit{Do Not: give up your coat.}}  

Do not let the host take your coat or any other belongings.  The host will stick your coat with thirty other coats, and later you will have to incorporate into your departure scheme: digging through a hill of coats.  Instead, find an out-of-the-way and easily-accessible location to stow your belongings (pack lightly).  There exist innumerable variations on this warning, for example, not agreeing to hold Mindy's emotional support chihuahua while she goes to grab a martini and \lq\lq will be right back\rq\rq.  Moreover, do not do anything to hinder your autonomy.

\subsubsection*{\textit{Do: appear happy.}}

Appear genuinely happy throughout the evening.  To attain the \lq\lq genuine\rq\rq\ look, try not to focus on your smile, lest you strain.  Instead, imagine something that truly makes you happy.  For example, your cat.  Think about your cat.  Remember that you will see him soon.  This tactic also helps in the event of panic.  Also, laugh.  Laugh at statements that are intended to be funny, and do not laugh at statements that are not intended to be funny.  If you are not sure whether a comment is intended to be funny, mimic those around you.  More tips for success are offered in \textit{Results}.  

\subsubsection*{\textit{Do: Flesh out your model upon arrival.}}

Once a decent crowd surrounds you, define the model parameters.  Identify exits, as exemplified in Figure~\ref{fig:floorplan}.  If you are on a low floor, determine the feasibility of a window option.  Identify the windows that open.  The bathroom is of high interest, as you have an excuse to be in there; bedrooms are higher risk.  If you are on a high floor, note routes to stairwells and fire escapes.

Next, define the hyper-parameters governing the optimal exit.  These parameters include the density of people at particular locations, the overall rate of flow from region to region, and music loudness - in case the volume might cover the sound of a rusted fire escape gate creeking open.  These parameters are time-varying; sample them intermittently throughout the evening. 

\subsection*{\textbf{3 \hspace{2mm} GFOOEOPQ Phase One: \textit{Interact}}}

Armed with the lay of the land, it is time to begin Phase One of estimation.  Set your initial conditions on state variables $x_a^i$, and search ranges for parameters $\bm{\theta}$.  This step involves harnessing whatever information you already have regarding anyone you know, and employing wild-but-educated guesses for everyone else.  

The cost function $C_{interact}$ is derived elsewhere~\cite{abarbanel2013predicting}, and is written as:
\begin{align}
\label{eq:costfunction}
  C_{interact}(\bm{x}(n),\bm{\theta}) &= \sum_{i = Person 1}^Q \Big\{\sum_{j=1}^J\sum_{l=1}^L \frac{R^{i,l}_m}{2}(y_l^i(j) - x_l^i(j))^2 \notag\\
&+ \sum_{n=1}^{N-1}\sum_{a=1}^D \frac{R^{i,a}_f}{2}\big(x_a^i(n+1) -
f_a^i(\bm{x}^i(n),\bm{\theta}))\big)^2\notag\\ &+ \sum_{a=1}^D \frac{R^{i,a}_s}{2}\big(x_a^i(0) - \hat{x_a^i}(0)\big)^2\Big\} + \lambda \frac{T-Q}{T}.
\end{align}
\noindent
The first term of Equation~\ref{eq:costfunction} is a standard least-squares measurement error.  Here, measurements $y_l^i$ of measurable state variables $x_l^i$ are taken at the timepoints $j$ of observations, and we assume no transfer function from variables to measurements.  The second term represents model error, where the summation on variables is taken over all quantities, measurable and unmeasurable.  It is in this way that information contained in measurements is propagated to the model to estimate unknowns.  The third term represents intial conditions.   Coefficients $R^{i,l}_m$, $R^{i,a}_f$, $R^{i,a}_s$ are inverse covariance matrices for each term, respectively, and their utility will become clear below in this Section.  The fourth term is an equality constraint that assigns greater reward as the fraction of state vectors sampled by the user increases: $\lambda$ is a Lagrange multiplier of the desired strength, Q is the number of people sampled, and T is the total number of people\footnote{The user can add complexity to this constraint as needed.  For example, to tailor to the extreme at small T, require that all T be sampled.  For large T, employ a reasonable maximum threshold.}.  The optimum of Equation~\ref{eq:costfunction} corresponds to a solution in terms of state and parameter estimates.

To predict the likelihood of a particular facial expression on a particular person ($x_a^i$) on Monday, you will take the estimate, write it into your model $\bm{f}$, and integrate to Monday.  In principle this is straightforward.  For a nonlinear model, however, multiple solutions will exist.  If estimates are poorly resolved, they may appear identical, and we will not know which to choose.

Figure~\ref{fig:faces} illustrates the potentially disastrous consequences of failing to resolve multiple solutions until it is too late.  It depicts a two-dimensional slice of the solution for four state vectors, where the axes are \textit{identity} and \textit{facial expression}.  The left-most column  corresponds to a poorly-resolved estimates for state vectors, which yield inconclusive predictions.  (In practice, often there will be a tip-off to the user that estimates are insufficiently resolved.  For example, you do not recall Aleks (State Vector 1) having three pairs of eyes.)  Columns 2-4 display results of annealing, for three distinct behavioral strategies.  The predictions now are clear: a distinctly welcoming greeting (Column 2), distinct ambiguity (Column 3), and a distinctly unwelcome greeting (Column 4), respectively. 

You shall achieve the resolution required for identifying the global optimum via simulated annealing.  This routine is an iterative procedure, where in Phase One the iteration occurs in the re-weighting of covariance matrices for the measurement versus model terms, $R_m^i$ and $R_f^i$.  First, $R_m^i$ is set to a constant and $R_f^i$ is defined as: $R_f^i = R_{f,0}^i\alpha^{\beta}$.  The annealing parameter $\beta$ is set to a low value so that $R_m^i \gg R_f^i$.  Relatively free from model constraints, the cost function surface is smooth and there exists one minimum of the variational problem that is consistent with the measurements.  Then you will iteratively increase $\beta$, gradually imposing model constraints and updating the cost estimate.  The aim is to remain sufficiently near to the lowest minimum at each iteration so as not to become trapped in a nearby minimum as the surface becomes resolved.

Now, at this point, you the Reader may be thinking: \textit{This sounds nice in principle - but in practice?  What fraction of people do I talk to?  How long do I talk?}  Indeed, the breadth of choice may seem overwhelming.  Further, there is a cost/benefit analysis to consider.  The longer you interact, the longer your baseline of measurements and hence the better-resolved your estimation.  Meanwhile, you aim to minimize this baseline and move onto disappearing.  As GFOOEOPQ users are promised no requirement of gut instinct, a rough guide is offered in the form of Table 1.  

Table 1 offers a template for about how much - and how little - to know about these people before moving to leave.  It is taken from one of the simulated models reported in \textit{Results}; see the caption for details.  When you are able to complete a similar table, consider yourself ready.  For comparison, two templates representing terrible performances are offered in \textit{Appendix}: these are cases in which: 1) insufficient data are gathered (the \lq\lq too little\rq\rq\ scenario), and 2) you are forcibly removed from the party prior to estimation (\lq\lq too much\rq\rq).  Note that in accordance with the Bayesian framework, there is a nonzero probability that any similarities of the simulated state vectors of Table~\ref{table1} to real people are purely coincidental. 
\end{multicols}
\setlength{\tabcolsep}{5pt}
\begin{table}[H]
\small
\centering
\begin{tabular}{p{2.5cm}|p{3.5cm}|p{6cm}|p{5cm}}
 \textbf{Name} & \textbf{Description} & \textbf{Proof that you're listening} & \textbf{Proof that you're watching} \\\midrule
  & \Fontskrivan{\Large{Graduate student in English}} & \Fontskrivan{\Large{Is working on a satire of Fyodor Dostoevsky's \lq\lq \textit{Crime and Punishment}\rq\rq, wherein Raskolnikov has no arms.}} & \Fontskrivan{\Large{Didn't shave today.}} \\\hline 
  \Fontskrivan{\Large{\textit{Freddy}}} & \Fontskrivan{\Large{Professor in mathematics}} & & \Fontskrivan{\Large{Removed his shoes at the door even though no one else did.}} \\\hline
 \Fontskrivan{\Large{\textit{Amir}}} & \Fontskrivan{\Large{Professor emeritus of neurophysiology}} & \Fontskrivan{\Large{Is amid his thirty-fourth year of developing a cure for ambition.}} & \Fontskrivan{\Large{Keeps flirting}}$\bm{^\nu}$ \Fontskrivan{\Large{with the spouse of the math professor and being way too obvious about it.}} \\\hline
 \Fontskrivan{\Large{\textit{Persephone?  No, that can't be right.  Pat?}}} & \Fontskrivan{\Large{Professor of physics}} & \Fontskrivan{\Large{Has published extensively on the theoretical danger of strings.}} & \Fontskrivan{\Large{Is pretty.}}\\\hline
 \Fontskrivan{\Large{\textit{Fahid}}} & \Fontskrivan{\Large{President of the university}} & \Fontskrivan{\Large{Engineer-turned-administrator.  Claims to have invented the stapler, but his timeline sounds fishy.}} &  \\\hline
 \Fontskrivan{\Large{\textit{Aleks}}} & \Fontskrivan{\Large{Dean of something}}  & \Fontskrivan{\Large{Academic-turned-administrator.  Is spearheading an unpopular proposal to require all science faculty to read a literary novel at least once per decade, and all humanities faculty to learn algebra.}} &  \\\hline
  & \Fontskrivan{\Large{Spouse of the math professor}} & \Fontskrivan{\Large{Also claims to have invented the stapler.}} & \Fontskrivan{\Large{Pocketed four macaroons off the dessert tray.}} \\\hline 
 \Fontskrivan{\Large{\textit{JiYoung}}} &  &  \Fontskrivan{\Large{Is allergic to bactrim.}}  &   \\\bottomrule
\end{tabular}
\caption{\textbf{A template representing an acceptable set of interactions at the party, in number and substance.  If you can complete a table like this, you are ready to proceed with estimation for Phase One}.  \textit{General notes}: This looks fine.  You met eight of a total of 20 people (all 20 not shown), a decent ratio.  You are listening (Column 3) and observing (Column 4) in reasonable doses.  ($\bm{^\nu}$Good eye on the flirting; body language can be hard to interpret.)  For instructive terrible templates, see \textit{Appendix}.  (Note: In accordance with the Bayesian framework, there is a nonzero probability that any similarities of the simulated state vectors of Table~\ref{table1} to real people are purely coincidental.)} 
\label{table1}
\end{table}
 \begin{multicols}{2}
\subsection*{\textbf{4 \hspace{2mm} GFOOEOPQ Phase Two: \textit{Exit}}}
You have almost made it, Dear Reader: it is time to leave!  In this phase, model component $\bm{f}_{me}$ becomes a function of Parameter $Exit$.  You will seek the value of $Exit$ that drives $\bm{f}_{me}$ to zero, or as close to zero as possible.  That is, \textit{you desire no interaction with any state vector during this phase}.  
\end{multicols}
.\begin{figure}[H]
\centering
  \includegraphics[width=0.99\textwidth]{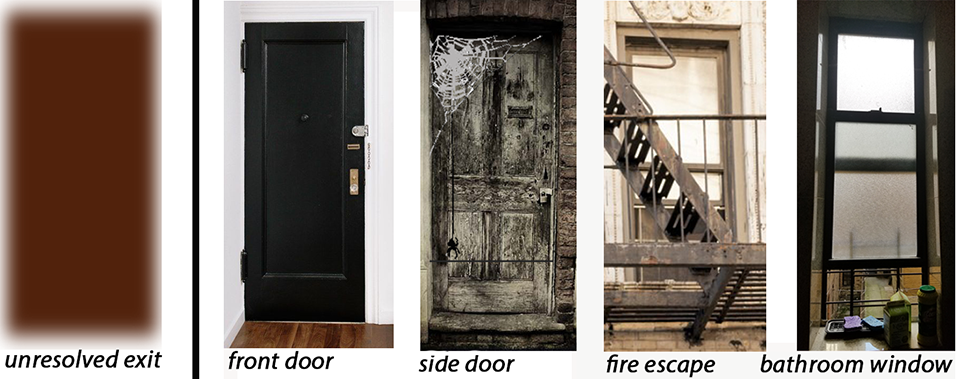}
  \caption{\textbf{Estimates of optimal exit parameters for unresolved (left) versus resolved (everything to the right) cases, illustrating the utility of annealing to avoid degenerate solutions.}  At left is the unresolved estimate obtained without the use of annealing: a blurry rectangular object.  To the right are resolved estimates obtained in simulations that employed annealing: front door, back door, fire escape, and window that opens onto the fire escape.}
  \label{fig:exits}
\end{figure}
\begin{multicols}{2}
As noted, the optimal exit depends on hyper-parameters $\bm{\zeta}$, which you have been sampling throughout Phase One.  Now you will map these $\bm{\zeta}$ to $Exit$ via some transfer function $\bm{g}$\footnote{Building the map $\bm{g}$ requires some dexterity of imagination.  For example, throughout Phase One you may note that the region by the front door is not typically crowded.  But the density of people is high and the rate of flow through the hall leading to the front door is high, so it will be hard to predict whether the front door's neighborhood will be crowded at any time.  Meanwhile, the fire escape is secluded, and the music is sufficiently loud to camouflage a gate creek.  On the other hand, in the event that you are seen leaving, Front Door is easier to explain than Fire Escape.  Moreover, the cost/benefit analysis is not trivial.  If you do your best here, GFOOEOPQ will do its best.}, and enter representative values into GFOOEOPQ.  The cost function is now written as $C_{exit}$:
\begin{align}
\label{eq:costfunction2}
  C_{exit}(\bm{x}(n),\bm{\zeta}) &= \sum_{i = Person 1}^Q \Big\{\sum_{j=1}^J\sum_{l=1}^L \frac{R^{i,l}_m}{2}(y_l^i(j) - x_l^i(j))^2 \notag\\
&+ \sum_{n=1}^{N-1}\sum_{a=1}^D \frac{R^{i,a}_f}{2}\big(x_a^i(n+1) \notag\\ & -
f_a^i(\bm{x}^i(n),Exit(\bm{g}(\bm{\zeta})))\big)^2\notag\\ &+ \sum_{a=1}^D \frac{R^{i,a}_s}{2}\big(x_a^i(0) - \hat{x_a^i}(0)\big)^2 \notag\\ &+ 10^7 \sum_{i = Person 1}^Q\sum_{j=1}^J\sum_{l=1}^L y_l^i(n)\Big\}.
\end{align}
\noindent
Equation~\ref{eq:costfunction2} possesses some key differences from Equation~\ref{eq:costfunction}.  Here, $\bm{f}$ evolves according to hyper-parameters $\bm{\zeta}$ rather than the $\bm{\theta}$ governing state vector evolution.  The initial conditions are set to their values at the final timepoint of the estimation window from Phase One.  And the main difference from Equation~\ref{eq:costfunction} is the treatment of measurements.  The measurement term is unchanged, but at the start of this estimation window, \textit{there must be no measurable state vectors}: the coast is clear.  To that end, the fourth term of Equation~\ref{eq:costfunction2} is an equality constraint that heavily penalizes - by a factor of 10,000,000\footnote{Cranking it higher introduces numerical issues.} - any measurement that occurs during the estimation window.  The optimal exit is the one that produces the fewest measurements within this window and simultaneously drives $\bm{f}_{me}$ to zero.  (Indeed, an optimal estimation during Phase Two requires that no measurements be made.  Ponder that at will.)  

Finally, due to the nonlinearity of the transfer function $\bm{g}$ that maps hyper-parameters $\bm{\zeta}$ to \textit{Exit}, you will again encounter degenerate solutions (Figure~\ref{fig:exits}).  Here, annealing is employed not in the relative weights of measurement-versus-model, but rather in the number of hyper-parameters: adding one at a time.  

A caution: you might expect Phase Two to be a breeze, as the social interactions are done.  Phase Two, however requires extreme dexterity.  Due to the ad-hoc nature of transfer function $\bm{g}$, your solution will not have predictive power for long.  You will have time to glance at the solution, glance over your shoulder, and then go for it or abort. 
\end{multicols}
\section{RESULTS}
\begin{multicols}{2}

\subsection*{\textbf{Simulations}}
I considered four simulated models, each defined by a number and constituency of people.  The numbers were 20, 31, 38, and 55, respectively.  Twenty was a lower limit chosen to avoid the slippery scenario in which so few people are present that your disappearance will be noted even if not directly witnessed (see \textit{Discussion}).  

For each model, in Phase One a global optimum was sought for three distinct predictions come Monday: a favorable disposition toward you, an unfavorable disposition, and a disposition that is uninformative or otherwise difficult to characterize (the \lq\lq ambiguous\rq\rq\ solutions of Figure~\ref{fig:faces}).  That is, three distinct recipes for the user's behavior were sought, the latter two for instructive comparison.  

For each of these twelve designs, 27 distinct choices were made regarding interaction strategy, including the fraction of people to approach, what to say to them, and how frequently to record their measurable features.  Each of these 324 experiments was run 10,000 times, each initialized with a different prior.  In Phase Two, again 10,000 trials were run for each model, over 43 distinct choices for the transfer function $\bm{g}$ between hyper-parameters and exit location.

Across all experiments, the percent convergence of paths ranged from 69 to 99 per cent, and estimations showed a roughly Gaussian distribution about their true values.  The illustrations in \textit{Methods} were taken from trials that yielded a minimum of 90 per cent convergence.  Simulations were run on a 720-core, 20-node CPU cluster over three weeks, or about 30,481,920 CPU hours.  The complete report is exhaustive (email if interested).  Here, the salient take-aways are distilled into handy tips. \\

\noindent
\textit{Tips for optimal interactions during Phase One}
\begin{itemize}
  \item Aim to sample 40 per cent of the population.  Above this fraction, the success rate begins to flatline, and below it the probability of obtaining the ambiguous facial expression on Monday rises as a power law in the fraction of people left unsampled. 
  \item In response to the question, \lq\lq How are you?\rq\rq, do not tell the person how you are.  Respond favorably and in one syllable. 
  \item Do not stare at people while working up the nerve to approach.  This appears creepy.  Make no eye contact until you are ready to say something.
  \item Do not borrow things with the intention of returning them later.  If you are observed, that intention is not the one that will be assumed.
  \item Do not eat a fraction of food that is large compared to the fraction of attendees you represent.
  \item Do eat some food, or carry some around on a plate.
  \item Stand in locations of high visibility to maximize your impact on anyone you do not directly approach.  Good places include: 1) the buffet table if it is not crowded, 2) some object of interest, for example, a fish tank, 3) a window with a view of something noteworthy, 4) a sofa (a bolder choice; be careful with this one).  
  \item Try to answer questions truthfully.  Lies are hard to remember.
  \item Try to be dryly witty.  Academics love dry wit.  Test potentially-dryly-witty comments beforehand on someone you trust who will rate them honestly.    
  \item If someone asks you where you got your degree, end the conversation politely.  You are never obligated to spend time talking to someone who asks people at a party where they got their degree.  
  \item If someone asks you a question that you don't hear and you feel that you have said, \lq\lq What?\rq\rq\ too many times: chuckle and say, "Yeah!"  The question was probably a yes-or-no question, and the answer is probably yes.
  \item Seem interested, but not too interested.  For examples of \lq\lq too interested\rq\rq, see Table~\ref{table3} of \textit{Appendix}. 
  \item If applicable: stick around for a speech.  If the setting is a workplace, a speech is likely.  The speaker will be either the most or least important person there, more likely the latter if much alcohol is present.  Even if you have completed your analog of Table~\ref{table1}, stay if you can stomach it.  It will be a good thing to prove that you remember come Monday.  It is also an opportunity for visibility: position yourself close to the speaker for maximum exposure.   
  \item Use your idiosyncracies as assets.  You don't have to try as hard as you think you do to seem normal.  This is academia: not a lot of normal people make it in.  So go ahead and tell them about your chestnut collection.  Their response might pleasantly surprise you. 
\end{itemize}
\subsection*{\textbf{Preliminary results at two real-life events}}

GFOOEOPQ was tested at two real events: a banquet at a dean's New Jersey residence (name and school are withheld to protect privacy), and a birthday party for the two-year-old daughter of an old college friend.  Both cases successfully identified a global optimum for interaction and exit phases.  The exit parameters corresponding to the global optima for Phase Two, however, were Chimney and Air Duct, respectively - indicating that GFOOEOPQ underestimates risk.  Unwilling to risk being correct in this suspicion, I instead left by the front door in both cases.  Fortunately, the attendees still appear positively disposed toward me, and if any witnessed my departure, they are keeping mum. 
\end{multicols}
\section{DISCUSSION}
\begin{multicols}{2}

While clearly there is honing to be done, preliminary results are encouraging.  In addition, an expansion of GFOOEOPQ is planned, in terms of an alternative formulation of the dynamical problem.  This is described here in qualitative terms. 

\subsection{Reverse formulation: sneak in late}

Rather than leaving as soon as possible, you might instead sneak in unseen as late as possible and stay until an acceptably-late hour.  This is a dicey enterprise.  It is worth considering, however, to permit GFOOEOPQ users more flexibility in scheduling. 

Consider three conditions under which the sneak-in-late option might be feasible.  In one case, you the Reader already have some familiarity with the party location and thus can define the exits.  In another, everyone inside is already drunk by the time you arrive and unlikely to notice you enter.  Of course, you might ask: \textit{But how can I know they are already drunk?}  This brings us to the best-case scenario, wherein you have a friend already inside whom you have enlisted as a spy to provide you with the environmental hyper-parameters near entrance locations.  In ruminations thus far, having a spy appears to be the most desirable prior for attempting the reverse formulation.

I caution you not to attempt this maneuver until it has been incorporated into GFOOEOPQ and tested properly.  In the event that you chance it on your own, however, here is some advice.  1a) Do not ring their buzzer.  Wait outside for someone to come along, and walk in after that person.  If that person is uncomfortable letting you in, wait for someone else.  Patience!  Repeat: do not ring their buzzer.  1b) You could ring somebody else's buzzer.  Maybe those people are expecting a package delivery and will buzz you in blind.  3) Do not take the elevator, no matter how high the floor.  The elevator has you cornered.  4) Unless you have a spy inside, enter by the front (or main) door.  Before entering, seek a place to stash your belongings.  A corner of the stairwell at the top floor by the roof entrance, for example; no one ever goes up there.  Then open the front door without knocking.  If anyone is standing right there, you can feign having just stepped out for some air.

\subsection{What if you are the only one who came?}

Finally, let us address the severely sticky situation that may be lurking in your mind, wherein you are the only one who shows up to the party.  Generally, if there are fewer than about twenty people, you simply cannot leave without someone noting your absence.  You are stuck.  This is not so terrible if there are nineteen others present.  But if it is just you and the host, it can be acutely painful.  And unfortunately, I see no exit.  (The \lq\lq emergency phone call\rq\rq\ is a well-known clich\'{e} and immediately suspect).  My only advice on this one is to exercise common sense before choosing to go.  Namely, predict all on your own whether others are likely to show - based on who is throwing the party.  You can do that one in your head, can't you?  It's not rocket science.
\end{multicols}\newpage
\section{ACKNOWLEDGMENTS}
\begin{multicols}{2}
Thank You to the organizers of the bimonthly Mother/Daughter Potluck Jamboree at Sedgwick Middle School in West Hartford, Connecticut, for provoking me to develop the skills underlying this algorithm at a young age.
\end{multicols}
\section{APPENDIX: Examples of unsuccessful social sampling}
For comparison to Table~\ref{table1} of \textit{Methods}, here are two terrible template strategies.  In Table~\ref{table2}, your data acquisition is acutely insufficient for proceeding to Phase One of estimation.  In Table~\ref{table3} you instead overshoot, and are removed from the party by law enforcement prior to estimation.
\setlength{\tabcolsep}{5pt}
\begin{table}[H]
\small
\centering
\begin{tabular}{p{2.5cm}|p{3.5cm}|p{6cm}|p{5cm}}
 \textbf{Name} & \textbf{Description} & \textbf{Proof that you're listening} & \textbf{Proof that you're watching} \\\midrule
  & \LobsterTwo{\large{Graduate student}} & \LobsterTwo{\large{Wrote \lq\lq \textit{Crime and Punishment}\rq\rq}}$\bm{^\nu}$. & \LobsterTwo{\large{Didn't shave or iron his pants or tuck in his shirt.}}$\bm{^\xi}$ \\\hline 
 & & & \LobsterTwo{\large{Is tall.}}$\bm{^\xi}$ \\\hline
 & & & \LobsterTwo{\large{Is tall.}}$\bm{^\xi}$ \\\hline
 & & \LobsterTwo{\large{Likes pretzels}}$\bm{^\delta}$. &   \\\hline
 $\bm{^\pi}$& \LobsterTwo{\large{President of the university}} &  & \LobsterTwo{\large{Is wearing a blue shirt.}}$\bm{^\xi}$  \\\hline
 \LobsterTwo{\large{\textit{Cupcake}}}$\bm{^\mu}$ &  &  & \LobsterTwo{\large{Is wearing a purple dress.}}$\bm{^\xi}$ \\\bottomrule
\end{tabular}
\caption{\textbf{A template representing insufficient effort at data-gathering at the party.  You are not ready to proceed with estimation.  Get back to work}.  \textit{General notes}:  The surplus of entries in Column 4 compared to 3 indicates that you are mainly observing, not interacting.  On the rare occasion in which you do interact, you are not really listening (see Note $\bm{^\nu}$ below).  This table reads as if you are filling it out from a distance, covertly staring at people from a corner.  Get out there and get busy.  \textit{Specifics}: $\bm{^\nu}$This person did not write \textit{Crime and Punishment}.  Fyodor Dostoevsky wrote \textit{Crime and Punishment} in 1866.  $\bm{^\xi}$You aren't making eye contact, are you?  Except for the shaving, your observations omit faces.  $\bm{^\delta}$This does not sound like a very substantive conversation.  Generally that is okay, but not if it is the only conversation you have.  $\bm{^\pi}$You don't know the President's name?  Seriously?  $\bm{^\mu}$Cupcake is unlikely to be this person's name.  You probably overheard someone else calling the person Cupcake, for example, a spouse.  This does not mean that you should use Cupcake.} 
\label{table2}
\end{table}
\setlength{\tabcolsep}{5pt}
\begin{table}[H]
\small
\centering
\begin{tabular}{p{2.5cm}|p{3.5cm}|p{6cm}|p{5cm}}
 \textbf{Name} & \textbf{Description} & \textbf{Proof that you're listening} & \textbf{Proof that you're watching} \\\midrule
  $\bm{^\nu}$\Fontauri{\large{Jameson Reilly; prefers \lq\lq Flipper\rq\rq)}} & \Fontauri{\large{Graduate student of English in his fourth year. Passed his advancement-to-candidacy exam \lq\lq by a hair\rq\rq.}} & \Fontauri{\large{Is working on a satire of Fyodor Dostoevsky's \lq\lq \textit{Crime and Punishment}\rq\rq, wherein Raskolnikov has no arms.  Worked for four months on the outline and is currently proofreading his second draft.  Has engaged in a collaboration with the theatre department on a musical comedy of \lq\lq \textit{Crime and Punishment}\rq\rq, wherein Raskolnikov has no arms but still manages to excel at tap dancing.  Is in negotiations with a gaming company on a choose-your-own-adventure video version of \lq\lq \textit{Crime and Punishment}\rq\rq\ set in modern times, wherein the user - i.e. Raskolnikov - can choose as his victim among the pawnbroker, Cable repair technician, orthodontist, and other possibilities that catch the user's fancy.  Flipper left off his proofreading today on page 324 to come to this party.  Is so focused on his manuscript that he didn't even bother allocating time to shave before he came.}} & \Fontauri{\large{Looks tired.  Generally I am all for ambition, but I'm a little worried that Flipper is spending too much time on his work and too little taking care of himself.  He should be getting regular exercise and eating properly.  It doesn't look like he's doing either of those things.  Ultimately, this will negatively impact his work performance.  I should speak to him about this.}}\\\hline 
 $\bm{^\xi}$\Fontauri{\large{\textit{JiYoung Han, ne\'{e} Tan}}} & \Fontauri{\large{Is a friend who came to help the host make Boeuf Bourguignon, a tricky dish to prepare, in part because the sauce has to be thickened just right, and JiYoung is skilled at the beurre mani\'{e} method, which calls for a specific ratio of butter to flour.}} & \Fontauri{\large{Works from home as a clinical regulatory affairs director for Weill Cornell Medical College / New York Presbyterian Hospital, mainly serving their Upper East Side location.  Lives in Washington Heights.  Will not tell me which street.  Is allergic to bactrim.  Discovered this at age six when she broke out in a rash after taking a sulfate-based medication.  Had to be hospitalized for one day.  Will not tell me where the rash was.}} & \Fontauri{\large{Says that if I don't leave her alone she is going to call the police.\newline \newline  Update:\newline Is calling the po \hspace{2mm}}}
 \\\bottomrule
\end{tabular}
\caption{\textbf{Another template representing unacceptable data-gathering, where \lq\lq too little\rq\rq\ is not your problem.  Rather, you are evicted from the premises before even getting to the estimation}.  Although you produced no estimation, we have no trouble predicting how well your interactions will go come Monday.  \textit{Specifics}: $\bm{^\nu}$How long did you talk to this guy?  It sounds like he was verbose and happy to have an ear at hand.  In addition, your worry suggests that you may have made a true human connection.  This is good.  But you got lucky in happening upon a person who evidently loves to talk about himself.  $\bm{^\xi}$There are many features of this row that beg comments, but let's limit the critique to this: Do not ask strangers for their home addresses.} 
\label{table3}
\end{table}
\bibliographystyle{unsrt}
\nocite{*}
\bibliography{refs}

 


\end{document}